\documentclass[10pt,letterpaper,twocolumn]{article} 

\usepackage{ol2}
\usepackage[draft]{hyperref}
\usepackage{amsmath}

\begin{document}

\twocolumn[ 

\title{Polarization spectroscopy of an excited state transition}

\author{Christopher Carr, Charles S. Adams and Kevin J. Weatherill}

\address{
$^1$Department of Physics, Durham University, South Road, Durham, DH1 3LE, UK
\\

$^*$Corresponding author: k.j.weatherill@durham.ac.uk}

\begin{abstract}We demonstrate polarization spectroscopy of an excited state transition in room temperature cesium vapor. An anisotropy induced by a circularly polarized pump beam on the D2 transition is observed using a weak probe on the 6P$_{3/2}$ $\rightarrow$ 7S$_{1/2}$ transition. When the D2 transition is saturated, a sub-natural linewidth feature is observed which theoretical modeling shows is enhanced by Doppler averaging. Polarization spectroscopy provides a simple modulation--free signal suitable for laser frequency stabilization to excited state transitions.\end{abstract}


 ]





Polarization spectroscopy \cite{weim76} is a widely used Doppler-free technique that can provide a robust and modulation-free signal to which a laser can be frequency stabilized \cite{pear02}. The technique has predominantly been used on strong optical transitions from the ground state of atomic vapors \cite{harr05} where optical pumping induces birefringence in the medium or where birefringence is due to saturation effects \cite{java10}. In addition to ground state transitions, excited state spectroscopy is of growing interest for applications such as the search for stable frequency references \cite{bret93}, Rydberg gases \cite{mour98} and their application to electro-optics \cite{moha08} and non-linear optics \cite{prit10}, state lifetime measurement \cite{shen08}, optical filtering \cite{bill95}, multi-photon laser cooling \cite{wu09}, frequency up-conversion \cite{meij06} and frequency stabilization \cite{abel09}. In this work we extend the use of polarization spectroscopy to excited state transitions. By probing an infrared excited state transition with a large dipole moment we observe significant absorption and spectra with a signal to noise ratio of over 2~$\times$~$10^3$.


A schematic of the experimental setup is shown in Fig.~1 (a). A circularly polarized 852~nm pump beam stabilized to the 6S$_{1/2}$, $F$=4 $\rightarrow$ 6P$_{3/2}$, $F'$=5 transition, passes through a Cs room temperature vapor cell. A counter-propagating linearly polarized 1470~nm probe beam is scanned across the 6P$_{3/2},~$$F'$$=5 \rightarrow $7S$_{1/2}$, $F''$=4 transition. The scan is calibrated using a wavemeter. The relevant atomic level structure is shown in Fig 1 (b). The circularly polarized pump drives $\sigma^+$ transitions and transfers population towards the $|F',m_F=F'\rangle$ state, inducing an anisotropy in the medium. On the excited state transition, the component of the linearly polarized probe which drives $\sigma^-$  transitions is preferentially absorbed because there are no $\sigma^+$ allowed transitions from the $|F',m_F=F'\rangle$ state, resulting in a change in polarization of the probe. The electric field of the probe after the cell is \cite{hugh09}
\begin{eqnarray}
\vec{E}\!\!&=&\!\!E_0 \, {\rm exp}\left(ik_+L-\frac{\alpha_+L}{2}\right)\frac{{\rm exp}(-i\phi)}{2}(\hat{x}+i\hat{y}) \nonumber \\
 &+&\!\!E_0 \, {\rm exp}\left(ik_-L-\frac{\alpha_-L}{2}\right)\frac{{\rm exp}(i\phi)}{2}(\hat{x}-i\hat{y}),
\end{eqnarray}
where the wavevectors $k_\pm=\frac{\omega}{c}n_\pm$,  $n_\pm$ are the refractive indices of the vapor for the circular polarized components which drive $\sigma^\pm$ transitions and $\alpha_\pm$ are the corresponding absorption coefficients. We also define $\Delta n = n^+-n^-$ and $\Delta \alpha=\alpha^+-\alpha^-$.

\begin{figure}[t]
\centerline{\includegraphics[width=7.5cm]{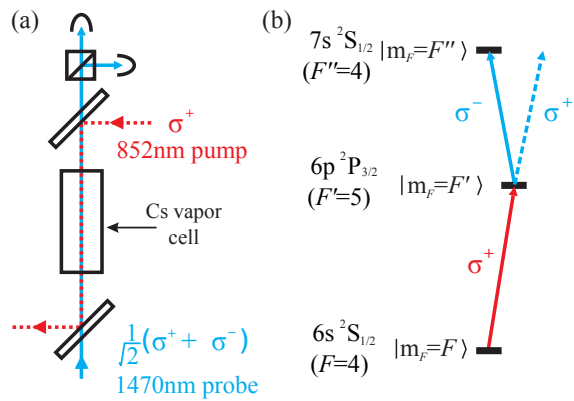}}
\caption{(color online) (a) Experimental setup. The 852~nm pump (1.6~mm 1/e$^2$ radius) and 1470~nm probe (1.2~mm 1/e$^2$ radius) beams counter-propagate through a 5~cm vapor cell. Probe rotation is measured by a polarimeter. (b) Energy level diagram. The circularly polarized pump drives $\sigma^+$ transitions and induces an anisotropy in the medium. The medium is probed with a linearly polarized laser on the excited state transition.}
\end{figure}

A polarizing beam splitting cube (PBS) oriented at angle $\phi = 45^{\circ}$ to the polarization vector of the probe resolves the probe electric field into orthogonal linear components which are detected using two Ge photodiodes. The resolved components $S_1$ and $S_2$ are proportional to $|\vec{E}|^2$ and are normalized so that the off-resonance transmission is one. Thus, $(S_1+S_2)/2\propto \Delta \alpha$ and the anisotropy $(S_1-S_2)/2 \propto \Delta n$.
\begin{figure}[t]
\centerline{{\includegraphics[width=7.5cm]{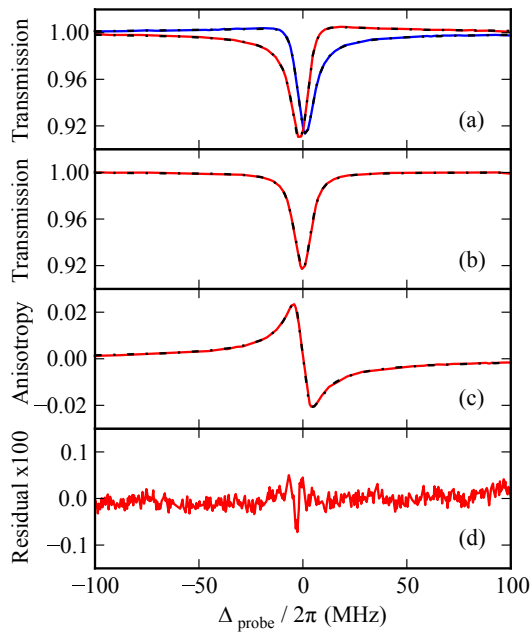}}}
\caption{(color online) Experimental spectra with theoretical fitting (dashed lines). (a) Individual signals $S_1$ (red line) and $S_2$ (blue line) recorded at the two photodiodes. (b) $(S_1+S_2)/2$ is a Lorentzian absorption profile and (c) $(S_1-S_2)/2$ is a dispersive shaped profile for the excited state transition. (d) The residual of the fit to the data shown in (c).}
\end{figure}
Fig.~2 shows spectra obtained for a moderate pump power of 500~$\mu$W and probe power of 10~$\mu$W. Fig.~2 (a) shows the individual spectra $S_1$ and $S_2$ recorded at the two photodiodes as a function of probe detuning while Fig.~2 (b) shows $(S_1+S_2)/2$ which is a Lorentzian profile with FWHM $\Gamma$ which can be written as,
\begin{equation}
\Delta \alpha = \frac{\Delta \alpha_0}{1+(2\Delta_{\rm probe}/ \Gamma)^2},
\end{equation}
where $\Delta_{\rm probe}$ is the detuning of the probe laser from resonance and $\Delta \alpha_0$ is the maximum difference in absorption at the line center. Fig.~2 (c) shows the polarization spectrum $(S_1-S_2)/2$. This signal is proportional to the dispersion described by,
\begin{equation}
\Delta n = \Delta \alpha_0\frac{2c}{\omega_0 \Gamma}\frac{\Delta_{\rm probe}}{1+(2\Delta_{\rm probe}/ \Gamma)^2}.
\end{equation}
The experimental data for $S_1$ and $S_2$ are fit using a combination of equations (2) and (3) yielding the linewidth $\Gamma$ of the feature. Fig.~2 (d) shows the residual to the fit of the data in Fig.~2 (c), the largest discrepancy (less than 0.1$\%$) arises from short-term fluctuations of the probe frequency at the two-photon resonance.

Fig.~3 shows the development of the polarization spectrum as a function of pump power. As the pump intensity is increased, the magnitude of the feature increases until the pump transition is saturated. The linewidth also increases with power broadening and eventually displays an Autler-Townes splitting \cite{flei05}. The resultant sub-feature at the centre of the main lineshape has the opposite gradient to the main signal and its magnitude and width also increase with pump intensity. Under these conditions of Autler-Townes splitting, it is no longer possible to model the data using a single Lorentzian and its concomitant dispersion, however an excellent fit is obtained using a pair of absorption/dispersion lineshapes.


\begin{figure}[t]
\centerline{\includegraphics[width=7.5cm]{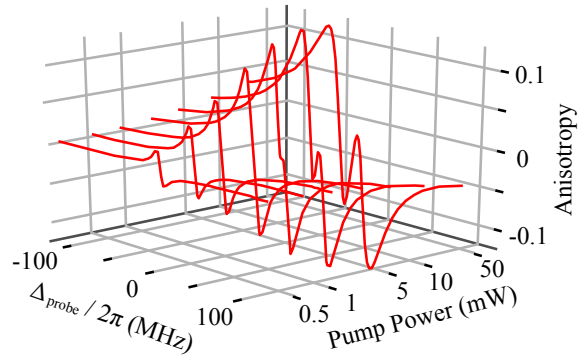}}
\caption{(color online) Experimental data. Evolution of the polarization spectra with increasing pump power.}
\end{figure}


The evolution of the on-resonance gradient and width of the features with pump power are shown in Fig.~4 (a) and (b) respectively. The on-resonance gradient increases with pump power before decreasing and changing sign with the appearance of the sub-feature at approximately 4.1~mW. Fig.~4 (b) shows the linewidth of the main and sub-features. At low pump powers, the main feature gives an excited state transition linewidth of 5.7~MHz. This linewidth increases with the square root of the pump power as expected. The linewidth of the sub-feature increases from zero also according to the square root of the pump power. Significantly, the linewidth of the sub-feature can be less than the natural linewidth of the 6P$_{3/2}$ (5.2~MHz) and 7S$_{1/2}$ (3.3~MHz) states.


To elucidate the origin of the sub-natural linewidth feature we solve the optical Bloch equations for the temporal evolution of the system including the component of the thermal velocity of the room-temperature atoms in the direction of the beams.  The density matrix approach provides a simplified description of the three-level system with the diagonal elements $\rho_{11}$, $\rho_{22}$ and $\rho_{33}$ being the populations of the 6S$_{1/2}$, 6P$_{3/2}$ and 7S$_{1/2}$ states respectively. The off-diagonal elements are the coherences between the states where the decay rates of the 7S$_{1/2}$ and 6P$_{3/2}$ states are 48.2~ns and 29.5~ns respectively \cite{theo84}. We consider the real part of the coherence between the intermediate and excited state Re$(\rho_{23})$ as it is proportional to the probe dispersion and thus our experimental signal. 

\begin{figure}[t]
\centerline{\includegraphics[width=7.5cm]{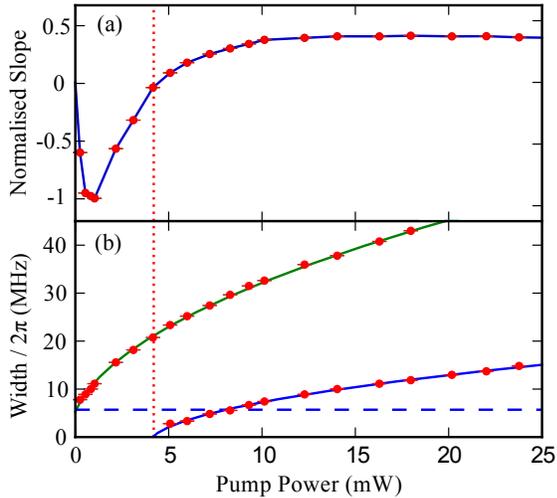}}
\caption{(color online) Analysis of dispersive lineshapes. (a) Measured gradient of the signal at $\Delta_{\rm probe}=0$ for varying pump power normalized to the maximum gradient. The solid line is a guide to the eye. The dotted vertical line indicates zero gradient. (b) Measured linewidth of the main feature and sub-feature with $\sqrt{P}$ scaling (solid lines). The dashed line indicates the natural linewidth of the excited state transition.}
\end{figure}

Fig.~5 shows the solution of the optical Bloch equations for Re$(\rho_{23})$ using Rabi frequencies $\Omega_{\rm pump}/2\pi=12.2$~MHz and $\Omega_{\rm probe}/2\pi=3.3$~MHz to match the average Rabi frequencies experienced by the atoms across the cell in the experiment. This simple model shows good qualitative agreement with the experimental data but quantitative agreement requires a full integration over the intensity profile of the beams as they are absorbed across the cell. An atom at rest has a dispersive feature with a positive gradient on resonance (red line). However, when velocity contributions are included (grey lines), the Doppler--averaged dispersive signal (blue line) has a sub-feature which is significantly narrower than the main feature. Thus the model shows that the onset of the narrow sub-feature is enhanced by the contributions of non-zero velocity classes and occurs at lower pump powers in thermal atoms than would be the case for cold atoms. Similar narrowing effects of thermal averaging have been observed in other systems \cite{ye02,bason09}.

The dispersive shaped feature from excited state polarization spectroscopy provides a convenient discriminant for laser frequency stabilization.
In future work we will apply this technique to the intermediate step of a three-photon excitation scheme \cite{vogt06} for use in Rydberg atom experiments \cite{prit10} and in the creation of ultra-cold ion and electron beams from laser-cooled atoms. \cite{hans06,clae05}

\begin{figure}[t]
\centerline{\includegraphics[width=7.5cm]{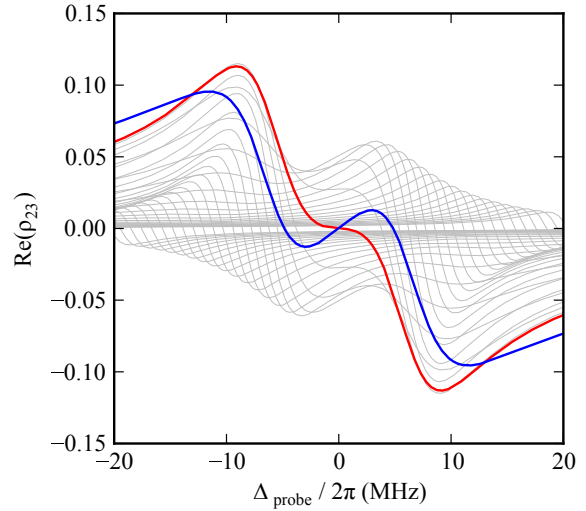}}
\caption{(color online) Theoretical modeling of lineshapes. The red lineshape is for zero velocity atoms. The blue line is the Doppler-averaged lineshape, multiplied by a factor of 4, for room temperature atoms. Each grey line represents an atomic velocity class separated by 2~m/s.}
\end{figure}

We thank Ifan Hughes for discussions and acknowledge support from Durham University and EPSRC.


\end{document}